\documentclass[12pt, a4paper]{article}
\pdfoutput=1
\usepackage{cite}
\usepackage{amsmath,amssymb}
\input{colordvi.tex}
\usepackage{color}
\usepackage{graphicx,hyperref}
\usepackage{comment}
\usepackage{multirow}

\hypersetup{
setpagesize=false,
 bookmarksnumbered=true,%
 bookmarksopen=true,%
 colorlinks=true,%
 linkcolor=red,
 urlcolor=blue,
 citecolor=blue,
}

\begin{document}

\begin{titlepage}

\begin{center}

\hfill UT-18-21

\vskip .75in

{\Large \bf 
Flavon Stabilization in Models with \\ \vspace{2mm} Discrete Flavor Symmetry
}

\vskip .75in

{\large
So Chigusa$^{(a)}$, Shinta Kasuya$^{(b)}$ and Kazunori Nakayama$^{(a,c)}$
}

\vskip 0.25in

$^{(a)}${\em Department of Physics, Faculty of Science,\\
The University of Tokyo,  Bunkyo-ku, Tokyo 113-0033, Japan}\\[.3em]
$^{(b)}${\em Department of Mathematics and Physics,\\
Kanagawa University, Kanagawa 259-1293, Japan}\\[.3em]
$^{(c)}${\em Kavli Institute for the Physics and Mathematics of the Universe (WPI),\\
The University of Tokyo,  Kashiwa, Chiba 277-8583, Japan}

\end{center}
\vskip .5in

\begin{abstract}

We propose a simple mechanism for stabilizing flavon fields with aligned
vacuum structure in models with discrete flavor symmetry. The basic idea
is that flavons are stabilized by the balance between the negative soft
mass and non-renormalizable terms in the potential.  We explicitly
discuss how our mechanism works in $A_4$ flavor model, and show that the
field content is significantly simplified.  It also works as a natural
solution to the cosmological domain wall problem.

\end{abstract}

\end{titlepage}


\renewcommand{\thepage}{\arabic{page}}
\setcounter{page}{1}
\renewcommand{\thefootnote}{\#\arabic{footnote}}
\setcounter{footnote}{0}

\section{Introduction}

There are many proposals that try to explain the observed pattern of
neutrino masses and mixings using discrete flavor symmetry (for reviews,
see
Refs.~\cite{Altarelli:2010gt,Ishimori:2010au,King:2014nza,King:2017guk}).
In these classes of models, the Lagrangian is assumed to be invariant
under some discrete symmetry and charged leptons and neutrinos obtain
masses through vacuum expectation values (VEVs) of scalar fields which
are charged under the discrete symmetry, called flavons.  Thus the
discrete symmetry is spontaneously broken by the VEVs of flavons.

Models with the spontaneous breakdown of discrete symmetry necessarily
have several discrete degenerate vacua. In general, such a model suffers
from the cosmological domain wall
problem~\cite{Vilenkin:2000jqa}. Domain wall problem in the context of
discrete flavor symmetry has been discussed in a few
papers~\cite{Riva:2010jm,Antusch:2013toa,Chigusa:2018hhl,King:2018fke}.
Ref.~\cite{Riva:2010jm} considered some possibilities that may avoid the
domain wall problem in the $A_4$ model based on the scalar potential
proposed in Ref.~\cite{Altarelli:2005yx}.  The inflation scale could be
lower than the flavor symmetry breaking scale so that the flavor
symmetry breaking occurs before inflation, but there is a flat direction
in the scalar potential in the model of Ref.~\cite{Altarelli:2005yx},
and hence this solution does not work unless the inflation scale is
lower than the soft supersymmetry (SUSY) breaking scale.  One
possibility is that (a part of) the discrete symmetry is explicitly
broken so that the VEV of flat direction does not produce exactly stable
domain walls.\footnote{
Present authors (S.C. and K.N.) have shown that the effect of quantum
anomaly under the color SU(3) is not enough to completely solve the
degeneracy of the discrete vacua at least in many of the known discrete
flavor models~\cite{Chigusa:2018hhl}.  }
Ref.~\cite{Riva:2010jm} also briefly discussed a possibility
that the flat direction in the scalar potential obtains a large field
value during/after inflation due to the Hubble mass correction, so that
domain walls are inflated away. However, we find out that actually it is
difficult to stabilize the flat direction in the desired way as far as
we work with the scalar sector of Ref.~\cite{Altarelli:2005yx}, because
the flat direction necessarily has unwanted higher dimensional operators
that is not suitable for the purpose of solving the domain wall problem.

In this paper, we propose a very simple alternative mechanism to
stabilize the flavon potential with desired alignment structure. In our
model, flavons are stabilized by the balance between the negative soft
SUSY breaking mass and non-renormalizable terms in the potential. The
field content is then significantly simplified, and we do not need any
additional field (so-called driving field). In addition, the
cosmological domain wall problem is naturally solved independent of the
inflation scale.

The structure of the paper is as follows.  In the next section, we
first review $A_4$ model briefly. In Sec.~\ref{sec:idea}, we explain the
basic idea of our novel mechanism of the flavon stabilization and study
a concrete model with $A_4$ symmetry in Sec.~\ref{sec:simple}. We show
how the domain wall problem is naturally avoided in our model in
Sec.~\ref{sec:DW}. Sec.~\ref{sec:conc} is devoted to our conclusions and
discussion.  $A_4$ representations are summarized in the Appendix.


\section{Brief description of $A_4$ model}

We assume the superpotential of the form
\begin{align}
	W = W_{\ell} + W_{\rm f},
\end{align}
where $W_{\rm f}$ is related to the stabilization of the flavon fields discussed in the next section, 
and $W_\ell$ is given by~\cite{Altarelli:2005yx} 
\begin{align}
 W_{\ell} &= \frac{y_e}{\Lambda} e^c H_d (\varphi_T \ell) +
 \frac{y_\mu}{\Lambda} \mu^c H_d (\varphi_T \ell)' +
 \frac{y_\tau}{\Lambda} \tau^c H_d (\varphi_T \ell)'' \notag \\
 &\quad + \frac{x_a}{\Lambda^2} H_u H_u \xi (\ell \ell) +
 \frac{x_b}{\Lambda^2} H_u H_u (\varphi_S \ell \ell) +
 \cdots,\label{Wl}
\end{align}
where $\ell$ denotes the $A_4$ triplet lepton doublets, while $A_4$
singlets $e^c, \mu^c, \tau^c, H_u$, and $H_d$ represent the right-handed
electron, muon, tau, the up-type Higgs, and the down-type Higgs,
respectively.  Also, $\varphi_T$, $\varphi_S$, and $\xi$ are flavon
fields and $\Lambda$ is the cutoff scale.
We assign the charges of the fields under the $A_4$ symmetry as in
Ref.~\cite{Altarelli:2005yx} (see Table~\ref{table:A4}).  Here
$(\cdots)$, $(\cdots)'$, and $(\cdots)''$ represent the contraction of
various $\textbf{3}$ representations that transform respectively as
$\textbf{1}$, $\textbf{1}'$, and $\textbf{1}''$.  In addition, the dots
in Eq.~\eqref{Wl} stand for the higher dimensional operators.  In this
paper, we take a basis of the triplet representation that diagonalizes
generators of the subgroup $Z_3 \subset A_4$.  For details of the
convention of the basis and the product of representations, see
Appendix~\ref{sec:appendix}.

If the flavon fields obtain VEVs of the following structure,
\begin{align}
	\left<\varphi_T\right> = (v_T,0,0),~~~\left<\varphi_S\right>=(v_S,v_S,v_S),~~~\left<\xi\right>=v_\xi,
	\label{align}
\end{align}
the diagonal mass matrix of the charged lepton is obtained, while the
neutrino mixing matrix becomes the so-called tri-bimaximal
form~\cite{Ma:2001dn,Harrison:2002er,Altarelli:2005yx}.  Here and below,
we use the same notation $\varphi_T$, $\varphi_S$, and $\xi$ for the
scalar component of the corresponding superfield.  The recent neutrino
oscillation data requires deviation from the tri-bimaximal
form. Extensions of the $A_4$ model to account for the observed data are
found, {\it e.g.} in Refs.~\cite{Shimizu:2011xg,Kang:2018txu}. For a
while, however, we study this minimal setup to avoid unnecessary
complexity and discuss modification related to the deviation from
the tri-bimaximal structure in Sec.~\ref{sec:other}.

\section{Basic idea for the novel simple flavon stabilization} \label{sec:idea}

\begin{table}
\begin{center}
\begin{tabular}{|c|cccc|cc|ccc|} 
\hline
    ~        &  $\ell$ & $e^c$ & $\mu^c$ & $\tau^c$ & $H_u$ & $H_d$ & $\varphi_{T}$ & $\varphi_{S}$ & $\xi$ \\ \hline
 $A_4$   & ${\bf 3}$ & ${\bf 1}$ & ${\bf 1''}$ & ${\bf 1'}$ & ${\bf 1}$ & ${\bf 1}$ & ${\bf 3}$& ${\bf 3}$& ${\bf 1}$    \\ 
 $Z_{12}$& $\rho^5$ & $\rho^7$ & $\rho^7$ & $\rho^7$ & $1$ & $1$ & $1$& $\rho^2$& $\rho^2$    \\ 
 U(1)$_R$ & $\frac{5}{6}$ & $\frac{5}{6}$ & $\frac{5}{6}$ & $\frac{5}{6}$ & 0 & 0 & $\frac{1}{3}$ & $\frac{1}{3}$ & $\frac{1}{3}$ \\
 \hline
\end{tabular}
\caption{Charge assignments under $A_4$, $Z_{12}$ and U(1)$_R$ for leptons
and various Higgs and flavon fields. Here $\rho \equiv e^{\pi i /6}$ is
an element of $Z_{12}$ rotation.}  \label{table:A4}
\end{center}
\end{table}

In the most known models so far, the flavon fields are stabilized at the
renormalizable level.  Then we need to introduce several driving fields
in addition to the flavon fields in order to obtain the alignment
structure of Eq.~\eqref{align}, and the resulting superpotential is
complicated~\cite{Altarelli:2005yx}.  Moreover, there may remain a flat
direction in the scalar potential that requires additional stabilization
mechanism and it makes domain wall problem serious.

We propose a simple superpotential that successfully realize the
alignment structure of Eq.~\eqref{align}.  The basic idea is that the
flavon fields are stabilized by the balance between the negative soft
SUSY breaking mass and non-renormalizable terms in the
potential. Schematically, we just assume
\begin{align}
	W_{\rm f} = \frac{\varphi^n}{n \Lambda^{n-3}},  \label{Wn}
\end{align}
where $\varphi$ collectively denotes the flavon fields. Giving the tachyonic soft 
SUSY breaking mass term as $V_{\rm SB} = -m_\varphi^2|\varphi|^2$, we can stabilize flavons at 
\begin{align}
	\left<\varphi\right> \sim \left(\frac{m_\varphi \Lambda^{n-3}}{n-1}\right)^{1/(n-2)},   \label{vphi}
\end{align}
up to the complex phase, which will be fixed after taking into account the supergravity correction.
We do not need any additional field to stabilize
the flavon.  This type of potential is considered in the context of
thermal inflation~\cite{Lyth:1995ka}.  Interestingly, this stabilization
mechanism can naturally solve the cosmological domain wall problem as
explained in Sec.~\ref{sec:DW}.  Of course, it is non-trivial whether or not we
can correctly obtain a desired alignment structure of Eq.~\eqref{align}
for this type of potential.  We will see that it is actually
possible with showing a concrete example in the next section.\footnote{
In the context of non-SUSY $S_4$ flavor model, 
a similar vacuum alignment mechanism with renormalizabe scalar potential 
was considered in Ref.~\cite{deMedeirosVarzielas:2017hen}.
}

Note that in this case, we have
\begin{align}
	\frac{\left<\varphi\right>}{\Lambda} \sim  \left(\frac{1}{n-1}\frac{m_\varphi}{\Lambda}\right)^{1/(n-2)}
	\simeq 
	\begin{cases}
	\displaystyle 5.8\times 10^{-4} \left( \frac{m_\varphi}{10^3\,{\rm TeV}} \right)^{1/2} 
	\left( \frac{10^{12}\,{\rm GeV}}{\Lambda} \right)^{1/2}~{\rm for}~n=4\\
	\displaystyle 6.7\times 10^{-3} \left( \frac{m_\varphi}{10\,{\rm TeV}} \right)^{1/4} 
	\left( \frac{10^{12}\,{\rm GeV}}{\Lambda} \right)^{1/4}~{\rm for}~n=6
	\end{cases},
\end{align}
where we show the numerical values for $n=4$ and $n=6$ as examples.  In
order for the tau Yukawa coupling $y_\tau$ to be within the perturbative
range, we need to have $\left<\varphi\right>/\Lambda \gtrsim 10^{-3} /
\cos\beta$ with $\tan\beta \equiv v_{H_u}/ v_{H_d}$ being the ratio of the VEVs of up- and down-type Higgs.
Thus, relatively small $\tan\beta$ $(\sim \mathcal O(1))$ may
be favored. For large $n$ (say, $n\gtrsim 6$) and/or large soft masses,
this inequality is rather easily satisfied. Neutrino masses are of the
order of
\begin{align}
	m_\nu \sim \frac{\left<\varphi\right>v_{H_u}^2}{\Lambda^2} \sim 3\times 10^{-2}\,{\rm eV}
	\left( \frac{10^{12}\,{\rm GeV}}{\Lambda} \right)\left( \frac{\left<\varphi\right>/\Lambda}{10^{-3}} \right)
	\sin^2\beta,
\end{align}
for $x_a$ and $x_b$ in Eq.~\eqref{Wl} being of order unity.

\section{Concrete $A_4$ model for the flavon scalar potential}   \label{sec:simple}

Now let us have a closer look at the flavon superpotential.  As a
concrete example, besides the $A_4$ symmetry, we impose an additional
$Z_{12}$ symmetry and the $R$-symmetry U(1)$_R$ in order to control the
flavon sector.  The charge assignments are summarized in
Table~\ref{table:A4}.  Actually, this model corresponds to the $n=6$
case described in the previous section, since lower dimensional
operators are forbidden by the charge assignments.

We further assume the superpotential for the flavon sector of the
following form:
\begin{align}
	W_{\rm f} = \frac{1}{6\Lambda^3}\left[ g_1(\varphi_T^2)^3 + g_2(\varphi_T^3)^2 
	+ g_3(\varphi_S^2)^3 + g_4(\varphi_S^2)^{''3} + g_5 \xi^6 \right].  \label{Wf}
\end{align}
Although we can also write down other terms of the sixth order in the
flavon fields allowed by the symmetry, their existence does not affect
the main results as far as coefficients $g_1,\dots, g_5$ in
Eq.~\eqref{Wf} are larger than those with other terms (say, by one order
of magnitude).  We will come back to this point in Sec.~\ref{sec:other},
and work with the superpotential Eq.~\eqref{Wf}.  In addition, flavons
are assumed to have soft SUSY breaking masses as
\begin{align}
	V_{\rm SB} = - m_T^2 \sum_{i=1}^3|\varphi_{Ti}|^2 - m_S^2 \sum_{i=1}^3|\varphi_{Si}|^2 - m_\xi^2|\xi|^2.
	\label{VSB}
\end{align}
The flavon scalar potential is then given by
\begin{align}
	V = \sum_{i=1}^3\left|\frac{\partial W_{\rm f}}{\partial \varphi_{Ti}}\right|^2
	+\sum_{i=1}^3\left|\frac{\partial W_{\rm f}}{\partial \varphi_{Si}}\right|^2
	+\left|\frac{\partial W_{\rm f}}{\partial \xi}\right|^2
	+ V_{\rm SB} + V_A.
\end{align}
Here we also include the so-called $A$-term potential $V_A$ induced by
the supergravity effect:
\begin{align}
	V_A = \frac{A}{2\Lambda^3}\left[ g_1(\varphi_T^2)^3 + g_2(\varphi_T^3)^2 
	+ g_3(\varphi_S^2)^3 + g_4(\varphi_S^2)^{''3} + g_5 \xi^6 \right] + {\rm h.c.},
\end{align}
where $|A| = m_{3/2}$ denotes the gravitino mass. For simplicity, we assume
that $m_{3/2}$ is smaller than the soft masses $m_{T}, m_S$, and $m_\xi$
and also that all coefficients $g_1,\dots, g_5$, and $A$ are real.\footnote{
	For example, $g_1$, $g_3$, $g_5$, and $A$ are taken to be real and positive without loss of generality, 
	but $g_2$ and $g_4$ are complex in general. The following discussion does not much depend on whether 
	$g_2$ and $g_4$ are real or complex.
}  It is easy to
see the stabilization of $\xi$, so we discuss the potential of
$\varphi_T$ and $\varphi_S$ below.

\subsection{Potential of $\varphi_T$}  \label{sec:phiT}

\begin{figure}
	\begin{center}
		\includegraphics[scale=0.5]{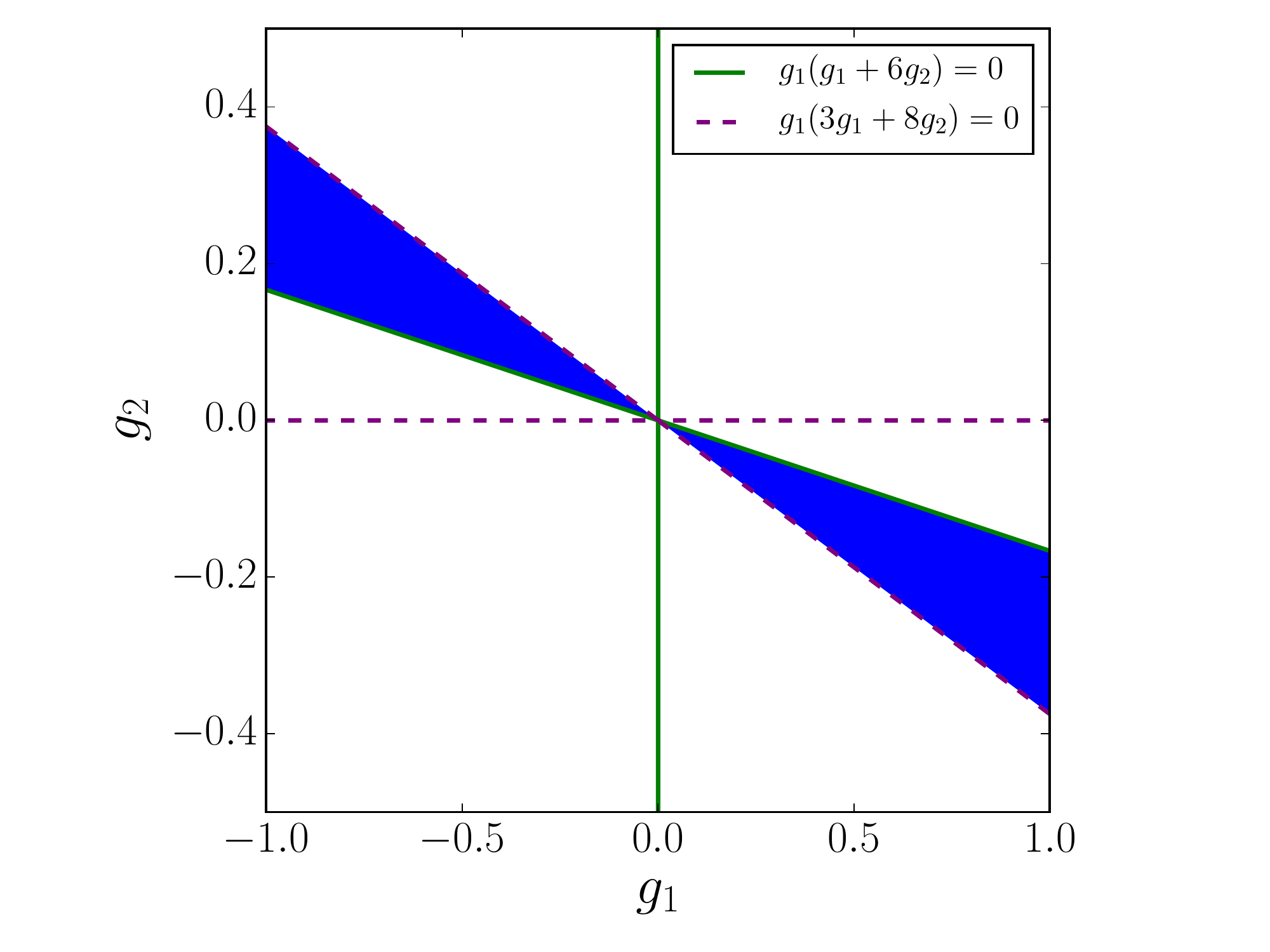}
		\caption{Parameter region consistent with the condition Eq.~\eqref{g1g2}. Blue region is allowed.\label{fig:phiT}}
	\end{center}
\end{figure}

First, in the $\varphi_T$ sector, we get a vacuum of the form of Eq.~\eqref{align} with
\begin{align}
	v_T = \left( \frac{1}{5} \right)^{1/8} \left( \frac{m_T \Lambda^3}{g_1+4g_2} \right)^{1/4}.
\end{align}
Of course, we can obtain other discrete vacua by transforming $\left<\varphi_T\right>$ 
with $A_4$ group element, but we pick this vacuum up.\footnote{
	As shown later in Sec.~\ref{sec:DW}, the flavons can naturally fall into  the desired minimum
	during inflation.
}
Here we ignore the $V_A$ term, which is justified as far as $m_{3/2} \ll m_T$.
Expanding the flavon fields around the vacuum as
\begin{align}
	\varphi_{T1} = v_T + \frac{1}{\sqrt 2}(\varphi_{T1}^R + i \varphi_{T1}^I),
	~~~\varphi_{T2} = \frac{1}{\sqrt 2}(\varphi_{T2}^R + i \varphi_{T2}^I),
	~~~\varphi_{T3} = \frac{1}{\sqrt 2}(\varphi_{T3}^R + i \varphi_{T3}^I),
\end{align}
we find out that $\varphi_{T1}^R$ obtains a positive mass-squared of $8m_T^2$.
On the other hand, $\varphi_{T2}^R$ and $\varphi_{T3}^R$ are mixed with each other. Their mass matrix is
\begin{align}
	\mathcal M_{R}^2=-\frac{4 m_T^2}{5(g_1+4g_2)^2}\begin{pmatrix}
		g_1^2+12g_1g_2 + 16g_2^2 & -g_1^2+16g_2^2\\
		-g_1^2+16g_2^2 & g_1^2+12g_1g_2 + 16g_2^2
	\end{pmatrix}.
\end{align}
The condition that both of them have positive mass eigenvalues is
\begin{align}
	g_1(g_1+6g_2)<0~~~{\rm and}~~~g_2(3g_1+8g_2)<0.  \label{g1g2}
\end{align}
Parameter regions consistent with this condition are shown in Fig.~\ref{fig:phiT}.
The same condition also ensures that $\varphi_{T2}^I$ and $\varphi_{T3}^I$ have
positive mass eigenvalues. All of them have mass-squared of $\mathcal
O(m_T^2)$.  $\varphi_{T1}^I$ obtains a mass only from the $V_A$ term
and its mass-squared is $\mathcal O(m_T m_{3/2})$, lighter than the
other scalar components, and has a positive sign for $g_1+4g_2 >
0$ $(g_1+4g_2  < 0)$ if $A > 0$ $(A<0)$. Therefore, we confirm that the $\varphi_T$ field is indeed
stabilized successfully at the vacuum of Eq.~\eqref{align}.

\subsection{Potential of $\varphi_S$}  \label{sec:phiS}

\begin{figure}
	\begin{center}
		\includegraphics[scale=0.5]{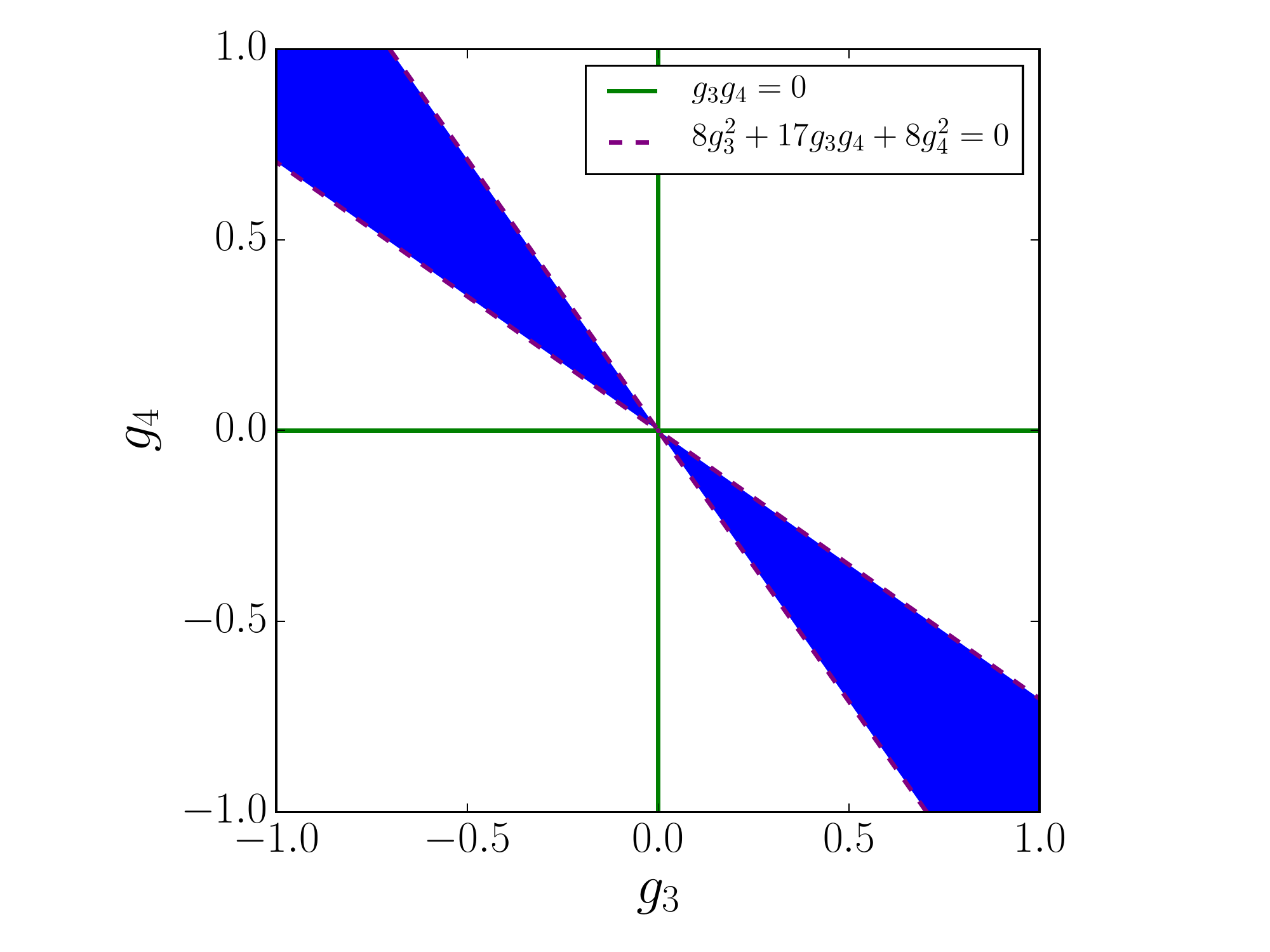}
		\caption{Parameter region consistent with the condition Eq.~\eqref{g3g4}. Blue region is allowed.\label{fig:phiS}}
	\end{center}
\end{figure}

Next, let us turn to the $\varphi_S$ sector. It is convenient to work
with the transformed basis:
\begin{align}
	\begin{pmatrix}
		\widetilde \varphi_{S1} \\ \widetilde \varphi_{S2} \\ \widetilde \varphi_{S3}
	\end{pmatrix}
	= U \begin{pmatrix}
		 \varphi_{S1} \\  \varphi_{S2} \\  \varphi_{S3}
	\end{pmatrix},~~~~~
	U \equiv \begin{pmatrix}
		1/\sqrt{3} & 1/\sqrt{3} & 1/\sqrt{3}  \\
		0 & -1/\sqrt{2} & 1/\sqrt{2} \\
		-\sqrt{2}/\sqrt{3} & 1/\sqrt{6} & 1/\sqrt{6}    
	\end{pmatrix}.
	\label{trans}
\end{align}
In this basis, the aligned vacuum in Eq.~\eqref{align} becomes
\begin{align}
	\left<\widetilde\varphi_S\right> = (\widetilde v_S, 0, 0).
\end{align}
We get a vacuum at
\begin{align}
	\widetilde v_S = \left( \frac{1}{5} \right)^{1/8} \left( \frac{m_S \Lambda^3}{g_3+g_4} \right)^{1/4}.
\end{align}
Again we ignore the $V_A$ term, which is justified as far as $m_{3/2} \ll m_S$.
Expanding the flavon fields around the vacuum as
\begin{align}
	\widetilde\varphi_{S1} = \widetilde v_S + \frac{1}{\sqrt 2}(\widetilde\varphi_{S1}^R + i \widetilde\varphi_{S1}^I),
	~~~\widetilde\varphi_{S2} = \frac{1}{\sqrt 2}(\widetilde\varphi_{S2}^R + i \widetilde\varphi_{S2}^I),
	~~~\widetilde\varphi_{S3} = \frac{1}{\sqrt 2}(\widetilde\varphi_{S3}^R + i \widetilde\varphi_{S3}^I),
\end{align}
we find out that $\widetilde\varphi_{S1}^R$ obtains a mass-squared of $8m_S^2$.
On the other hand, $\varphi_{S2}^R$ and $\varphi_{S3}^R$ are mixed with each other. Their mass matrix is
\begin{align}
	\mathcal M_{R}^2=-\frac{m_S^2}{5(g_3+g_4)^2}\begin{pmatrix}
		8g_3^2+13g_3g_4 + 2g_4^2 & 2\sqrt{3}(g_3+g_4)g_4\\
		2\sqrt{3}(g_3+g_4)g_4 & 3g_4(3g_3+2g_4)
	\end{pmatrix}.
\end{align}
The condition that both of them have positive mass eigenvalues is
\begin{align}
	g_3g_4 < 0~~~{\rm and}~~~8g_3^2+17g_3g_4 + 8g_4^2 < 0.  \label{g3g4}
\end{align}
Parameter regions consistent with this condition are shown in Fig.~\ref{fig:phiS}.
The same condition also ensures that $\varphi_{S2}^I$ and $\varphi_{S3}^I$ have
positive mass-squared eigenvalues. All of them have masses of $\mathcal
O(m_S^2)$.  The only exception is $\varphi_{S1}^I$, which obtains a mass
only from the $V_A$ term.  Its mass-squared is $\mathcal O(m_S
m_{3/2})$, lighter than the other scalar components, and has a positive
sign if $g_3+g_4 > 0$ $(g_3+g_4 < 0)$ if $A > 0$ $(A < 0)$. 
Thus, the $\varphi_S$ field is successfully stabilized at the vacuum of Eq.~\eqref{align}.

\subsection{Fermionic components}

So far we obtained the desired aligned vacuum of Eq.~\eqref{align}.  We
now discuss the mass matrix of the fermionic components of the flavons
$\varphi_T$ and $\widetilde \varphi_S$, which we call flavinos denoted by
$\psi_T$ and $\widetilde \psi_S$, respectively. Flavinos become massive
after flavons obtain VEVs.  For the $\varphi_T$ sector, the flavino mass
matrix is given by
\begin{align}
	\mathcal M_{\psi_T}=-\frac{m_T}{\sqrt{5}(g_1+4g_2)}\begin{pmatrix}
		5(g_1+4g_2) & 0 & 0\\
		0 & 0 & g_1-4g_2 \\
		0 & g_1-4g_2 & 0
	\end{pmatrix}.
\end{align}
For the $\widetilde \varphi_S$ sector, on the other hand, we have
\begin{align}
	\mathcal M_{\widetilde \psi_S}=-\frac{m_S}{\sqrt{5}(g_3+g_4)}\begin{pmatrix}
		5(g_3+g_4) & 0 & 0\\
		0 & -g_3+g_4/2 & -\sqrt{3}g_4/2 \\
		0 & -\sqrt{3}g_4/2 & g_3-g_4/2
	\end{pmatrix}.
\end{align}
Thus, all of them have masses of the order of the soft SUSY breaking
mass scale.

\subsection{Effects of other terms}   \label{sec:other}

We neglect many terms that are allowed by our charge assignments in the
flavon superpotential Eq.~\eqref{Wf}.  Thanks to the $Z_{12}$ and
U(1)$_R$ charges, the $\varphi_T$ sector, which gives the diagonal mass
matrix of the charged leptons, is completely decoupled from the
$\varphi_S$ and $\xi$ sector, which makes the tri-bimaximal form of the
neutrino mixing matrix.  There are eleven possible contractions of
$\varphi_T^6$ that transform as $\textbf{1}$ as a whole, while there are
seventeen possible contractions for the $\varphi_S$ and $\xi$ sector.

The effect of these terms can be estimated using some symmetry argument
as follows. First, we focus on the $\varphi_T$ sector.  If we refer to
one of the diagonalized elements of $A_4$, {\it e.g.} $T$ described in
Appendix~\ref{sec:appendix}, we can see that each component of the
flavon transforms as $\{\varphi_{T1}, \varphi_{T2}, \varphi_{T3}\} \to
\{ \varphi_{T1}, \omega \varphi_{T2}, \omega^2 \varphi_{T3} \}$ under
$T$, where $\omega \equiv e^{2\pi i/3}$.  Thus, the $\varphi_T$ sector
of the most general flavon superpotential in our model, which should be
invariant under $A_4$, takes the form of
\begin{align}
 W_{\rm f}^{(T)} \sim \frac{1}{\Lambda^3}\left(\varphi_{T1}^6 + \varphi_{T1}^4 \varphi_{T2}
 \varphi_{T3} + \cdots\right),\label{eq:Wft}
\end{align}
where the dots denote terms with more than two powers of
$\{\varphi_{T2}, \varphi_{T3}\}$.  Here, we omit all the
$\mathcal{O}(1)$ coefficients for simplicity.  It is straightforward
from Eq.~\eqref{eq:Wft} to see that $\varphi_T = (v_T, 0, 0)$ always
expresses one of the extrema of the potential with a properly chosen
value of $v_T$ thanks to the absence of terms linear in $\{\varphi_{T2},
\varphi_{T3}\}$.  Regarding the curvature of the potential around the
extremum, it is non-trivial to see whether the point we are considering
is a local minimum, a local maximum, or a saddle point.  However, we have already shown in Sec.~\ref{sec:phiT} that the
potential has a minimum at $\varphi_T = (v_T, 0, 0)$ for some parameter spaces spanned by $g_1$ and $g_2$.
Therefore, our flavon stabilization mechanism works well at least in the vicinity of such a parameter region.

Next, we consider the $\varphi_S$ and $\xi$ sector.  Again, it is
convenient to move to the basis expressed in Eq.~\eqref{trans}. In this
basis, one of the generators of $A_4$, which is denoted as $S$ in
Appendix~\ref{sec:appendix}, becomes diagonal:
\begin{align}
 \widetilde{S} \equiv U S U^{-1} = \left(
 \begin{array}{ccc}
  1 & 0 & 0 \\
  0 & -1 & 0 \\
  0 & 0 & -1 \\
 \end{array}
 \right).
\end{align}
Therefore, the $\varphi_S$ and $\xi$ sector of the most general flavon
superpotential has the form of
\begin{align}
 W_{\rm f}^{(S)} \sim \frac{1}{\Lambda^3}\left(\widetilde{\varphi}_{S1}^6 +
 \widetilde{\varphi}_{S1}^5 \xi + \widetilde{\varphi}_{S1}^4
 (\widetilde{\varphi}_{S2}^2 + \widetilde{\varphi}_{S2}
 \widetilde{\varphi}_{S3} + \widetilde{\varphi}_{S3}^2 + \xi^2) +
 \cdots\right),
\end{align}
where the dots represent terms with more than two powers of
$\{\widetilde{\varphi}_{S2}, \widetilde{\varphi}_{S3}, \xi \}$.  Again,
we omit all $\mathcal{O}(1)$ coefficients.  It can be seen that the same
argument as for the $\varphi_T$ sector follows, {\it i.e.},
$\widetilde\varphi_S=(\widetilde v_S,0,0)$ always remains an extremum of
the potential and hence our flavon stabilization mechanism works well in
some parameter regions at least in the vicinity of those shown in
Sec.~\ref{sec:phiS}.

We would also like to explain the deviation from the tri-bimaximal form
of the observed neutrino mixing matrix.  For this purpose, we follow the
procedure described in Refs.~\cite{Shimizu:2011xg, Kang:2018txu} and
consider a new flavon $\xi'$ that has charges $\{\textbf{1}', \rho^2,
1/3\}$ under our set up.  Again, we can find out an example parameter region that
stabilizes the flavon potential: we should just add a term proportional
to $\xi^{'6}$ to the superpotential Eq.~\eqref{Wn} and a term
$-m_{\xi'}^2 | \xi^{'}|^2$ to the SUSY breaking potential
Eq.~\eqref{VSB}.  In this simple example, $\xi'$ is completely decoupled
from other fields in the scalar potential and the minimization condition
of the potential remains unchanged except for the further condition for
the $\xi'$ VEV.  When we consider other possible contractions in the
superpotential, the number of allowed terms consist of $\{ \varphi_S,
\xi, \xi' \}$ is increased to twenty-nine.  However, the above
discussion still holds and there exists a finite volume region of the
parameter space that is consistent with our scenario.

\section{Solution to the domain wall problem}  \label{sec:DW}

Although the flavor structure can be well described by the discrete flavor symmetry with proper VEV alignment
of flavon fields, the domain wall problem arises if the spontaneous breaking of the flavor symmetry occurs after inflation,
because there are many degenerate minima of the flavon potential. In this section, we show that our mechanism of the 
flavon stabilization described in Sec.~\ref{sec:simple} results in a simple and natural 
description of the cosmological dynamics of the flavon fields that does not lead to the formation of domain walls.

The cosmological flavon dynamics depends on the soft mass $m_{\rm
soft}$, which collectively denote the soft SUSY breaking masses of
flavons, and the Hubble parameter $H$.  If $m_{\rm soft} \gtrsim H$
during inflation, flavons fall into one of the potential minima
dynamically and such a region expands that covers the whole
observable Universe without domain walls.

If $m_{\rm soft} \lesssim H$ during inflation, on the other hand, we need to include the negative Hubble-induced mass term
during and after inflation as~\cite{Dine:1995uk,Dine:1995kz}
\begin{align}
	V_{\rm SB} = - c_T H^2 \sum_{i=1}^3|\varphi_{Ti}|^2 - c_S H^2 \sum_{i=1}^3|\varphi_{Si}|^2 - c_\xi H^2|\xi|^2,
\end{align}
where $c_T$, $c_S$, and $c_\xi$ are positive constants of order unity.
These terms may come from the Planck-suppressed coupling among the inflaton and flavon fields 
in the K\"ahler potential. 
Since these negative Hubble-induced mass terms have the same form as the
negative soft mass terms Eq.~\eqref{VSB},
the structure of the minima is the same as those discussed in Sec.~\ref{sec:simple} with $m_{\rm soft}$ 
being replaced by $H$, which is typically much larger than  $m_{\rm soft}$.
Then, during the inflation, the flavons will settle down into the minimum of the potential at the VEVs, 
\begin{align}
	\left<\varphi_T\right> = (v^H_T,0,0),~~~\left<\varphi_S\right>=(v^H_S,v^H_S,v^H_S),
	~~~\left<\xi\right>=v^H_\xi,
	\label{alignH}
\end{align}
where
\begin{align}
	v^H_T, v^H_S, v^H_\xi \sim \left(\frac{H \Lambda^{n-3}}{n-1}\right)^{1/(n-2)},
\end{align}
since the flavons have masses of $\mathcal O(H)$ around this temporal minimum.\footnote{
	As explained in Sec.~\ref{sec:simple}, there are several light (pseudo) scalar fields 
	$\varphi_{T1}^I, \widetilde\varphi_{S1}^I$, and also $\xi^I$. Unless there is Hubble-induced 
	$A$-term~\cite{Kasuya:2008xp}, their mass-squared are of the order of $m_{3/2} H$ in the early 
	Universe and hence much lighter than other flavons. Although they obtain long wavelength isocurvature 
	fluctuations during inflation, they leave basically no observable signatures unless they are 
	directly related to the dark matter production or baryon asymmetry of the Universe.
}
Domain walls are thus inflated away no matter what happens before inflation.

After inflation, as the Hubble parameter decreases, the VEV of the temporal minimum Eq.~\eqref{alignH}
becomes smaller, and we expect that the flavons will track the temporal minimum. In order for this to happen, 
there is a lower bound on $n$ in Eq.~\eqref{Wn} depending on the background equation of state of the 
Universe as~\cite{Ema:2015dza}
\begin{align}
	n \geq \frac{6p}{3p-1} = \begin{cases}
		4 & {\rm for}~p=2/3, \\
		6 & {\rm for}~p=1/2,
	\end{cases} \label{n>p}
\end{align}
where $p$ is defined as the power law exponent of the cosmic scale factor: $a(t) \propto t^p$, 
hence $p=2/3$ $(1/2)$ for the matter (radiation) dominated Universe. If this condition is satisfied, 
flavons smoothly relax to the true potential minimum as $H$ becomes smaller than $m_{\rm soft}$. 
Otherwise, the flavon oscillations around the temporal minimum may be relatively enhanced as the Universe expands 
and flavons may overshoot the origin of the potential~\cite{Ema:2015dza}, which makes the whole flavon 
dynamics much more complicated and the domain wall formation may occur.

Since our model Eq.~\eqref{Wf} corresponds to the $n=6$ case, the flavon fields {\it do} track the temporal minima
of Eq.~\eqref{alignH}, and no symmetry restoration occurs after inflation if the inflaton effectively behaves as 
matter or radiation before the completion of reheating after inflation. 
Therefore, there are no domain walls in the observable Universe. This is a simple way to solve the domain 
wall problem in models with discrete flavor symmetry.

\section{Conclusions and discussion} \label{sec:conc}

We have proposed a simple mechanism to stabilize the flavon potential in
the model with discrete flavor symmetry.  It is achieved by the balance
between the negative soft mass and non-renormalizable terms with
appropriate charge assignments of the flavon fields. The flavon sector
is significantly simplified and the domain wall problem is naturally
solved. Although we have focused on the $A_4$ model as a concrete setup,
a similar mechanism should also work in models with other discrete
symmetries in general.  We leave this issue as future work.

Finally, we briefly discuss the cosmological aspects of this model.
Most flavons have masses of $m_{\rm soft}$ around the vacuum and some
flavons may be as light as $\sqrt{m_{3/2} m_{\rm soft}}$ if $m_{3/2}
\lesssim m_{\rm soft}$ as shown in Sec.~\ref{sec:simple}.  Flavons in
the $\varphi_T$ sector eventually decay into a Higgs boson plus 2 leptons
(and possibly their superpartners depending on the mass spectrum), and
its lifetime is typically much shorter than 1\,sec, which does not
affect the Big-Bang nucleosynthesis.  Flavons in the $\varphi_S$ and
$\xi$ sector decay into 2 Higgs bosons plus 2 leptons (and possibly their
superpartners). Since their decay rate is suppressed by $\Lambda^4$, the
lifetime can be much longer than 1\,sec for $m_S \ll 1000\,$TeV and
$\Lambda\sim 10^{12}\,$GeV and may affect Big-Bang nucleosynthesis.  In
such a case, we may need relatively low reheating temperature to avoid
the overproduction of flavons, which are dominantly in the form of
coherent oscillation.\footnote{ Even for the cosmological scenario
described in Sec.~\ref{sec:DW}, there is a finite amount of coherent
oscillation around the potential minimum~\cite{Ema:2015dza}, though the
oscillation amplitude is suppressed for $n$ satisfying inequality of Eq.~\eqref{n>p}.
}  Flavinos, the fermionic superpartners of the flavons, also have
masses of $\sim m_{\rm soft}$.  For $m_{3/2} \lesssim m_{\rm soft}$,
flavinos may decay into the gravitino and it can be a dominant source of
the gravitino production depending on the reheating temperature. The
gravitino decays into lighter SUSY particles unless the gravitino is the
lightest, and the final dark matter abundance is determined either by the
gravitino decay or thermal production, as usual.

\section*{Acknowledgments}

This work was supported by the Grant-in-Aid for Scientific Research C (No.\
18K03609 [KN]), and Innovative Areas (No.\ 26104009 [KN], No.\
15H05888 [KN], No.\ 17H06359 [KN]).  This work was also supported by
JSPS KAKENHI Grant (No.\ 17J00813 [SC]).

\appendix
\section{Notes on $A_4$ representations} \label{sec:appendix}

It is known that all the elements of the $A_4$ group can be written as
products of two elements $S$ and $T$, which are the generators of the
subgroup $Z_2$ and $Z_3$, respectively.  $A_4$ has a triplet
representation $\textbf{3}$ and three singlet representations
$\textbf{1}$, $\textbf{1}'$, and $\textbf{1}''$.  We adopt a basis of
the triplet representation that diagonalizes $T$.  In this basis, all
the elements of $A_4$ are represented as
\begin{align}
	&1 = \begin{pmatrix}
		1 & 0 & 0 \\
		0 & 1 & 0 \\
		0 & 0 & 1
	\end{pmatrix},~~~
	T= \begin{pmatrix}
		1 & 0 & 0 \\
		0 & \omega & 0 \\
		0 & 0 & \omega^2
	\end{pmatrix},~~~
	T^2=\begin{pmatrix}
		1 & 0 & 0 \\
		0 & \omega^2 & 0 \\
		0 & 0 & \omega
	\end{pmatrix}, \\
	&S = \frac{1}{3}\begin{pmatrix}
		-1 & 2 & 2 \\
		2 & -1 & 2 \\
		2 & 2 & -1
	\end{pmatrix},~
	ST= \frac{1}{3}\begin{pmatrix}
		-1 & 2\omega & 2\omega^2 \\
		2 & -\omega & 2\omega^2 \\
		2 & 2\omega & -\omega^2
	\end{pmatrix},~
	ST^2=\frac{1}{3}\begin{pmatrix}
		-1 & 2\omega^2 & 2\omega \\
		2 & -\omega^2 & 2\omega \\
		2 & 2\omega^2 & -\omega
	\end{pmatrix},\\
	&TS = \frac{1}{3}\begin{pmatrix}
		-1 & 2 & 2 \\
		2\omega & -\omega & 2\omega \\
		2\omega^2 & 2\omega^2 & -\omega^2
	\end{pmatrix},~
	TST= \frac{1}{3}\begin{pmatrix}
		-1 & 2\omega & 2\omega^2 \\
		2\omega & -\omega^2 & 2 \\
		2\omega^2 & 2 & -\omega
	\end{pmatrix},~
	TST^2=\frac{1}{3}\begin{pmatrix}
		-1 & 2\omega^2 & 2\omega \\
		2\omega & -1 & 2\omega^2 \\
		2\omega^2 & 2\omega & -1
	\end{pmatrix},\\
	&T^2S = \frac{1}{3}\begin{pmatrix}
		-1 & 2 & 2 \\
		2\omega^2 & -\omega^2 & 2\omega^2 \\
		2\omega & 2\omega & -\omega
	\end{pmatrix},~
	T^2ST= \frac{1}{3}\begin{pmatrix}
		-1 & 2\omega & 2\omega^2 \\
		2\omega^2 & -1 & 2\omega \\
		2\omega & 2\omega^2 & -1
	\end{pmatrix},~
	T^2ST^2=\frac{1}{3}\begin{pmatrix}
		-1 & 2\omega^2 & 2\omega \\
		2\omega^2 & -\omega & 2 \\
		2\omega & 2 & -\omega^2
	\end{pmatrix},
\end{align}
where we define $\omega \equiv e^{2 \pi i / 3}$.  The product of two
triplets are decomposed as $\textbf{3} \times \textbf{3} = \textbf{1} +
\textbf{1}' + \textbf{1}'' + \textbf{3}_S + \textbf{3}_A$.  For each
contraction, we use the convention of the coefficient as follows:
\begin{align}
	&{\bf 1} \sim a_1b_1 + a_2b_3 + a_3b_2 \equiv (ab),\\
	&{\bf 1}'\sim a_3b_3 + a_1b_2 + a_2b_1 \equiv (ab)',\\
	&{\bf 1}''\sim a_2b_2 + a_3b_1 + a_1b_3 \equiv (ab)'',\\
	&{\bf 3}_S \sim \begin{pmatrix}
		2a_1b_1-a_2b_3-a_3b_2 \\
		2a_3b_3-a_1b_2-a_2b_1 \\
		2a_2b_2-a_3b_1-a_1b_3
	\end{pmatrix}
	\equiv (ab)_{\bf 3_S},~~~
	{\bf 3}_A \sim \begin{pmatrix}
		a_2b_3-a_3b_2 \\
		a_1b_2-a_2b_1 \\
		a_3b_1-a_1b_3
	\end{pmatrix}
	\equiv (ab)_{\bf 3_A},
\end{align}
where $a=(a_1, a_2, a_3)$ and $b=(b_1, b_2, b_3)$ are triplets.

For the product of the same triplet $\varphi=(\varphi_1,\varphi_2,\varphi_3)$, we have
\begin{align}
	&(\varphi^2) = \varphi_1^2+2\varphi_2\varphi_3,~~~
	(\varphi^2)' = \varphi_3^2+2\varphi_1\varphi_2,~~~
	(\varphi^2)'' = \varphi_2^2+2\varphi_3\varphi_1, \\
	&(\varphi^2)_{\bf 3_S} = 2\begin{pmatrix}
		\varphi_1^2-\varphi_2\varphi_3 \\
		\varphi_3^2-\varphi_1\varphi_2 \\
		\varphi_2^2-\varphi_3\varphi_1
	\end{pmatrix},~~~
	(\varphi^2)_{\bf 3_A} = 0,\\
	&(\varphi^3)_{\bf 1}=2(\varphi_1^3+\varphi_2^3+\varphi_3^3-3\varphi_1 \varphi_2 \varphi_3).
\end{align}
In terms of the transformed basis $\widetilde \varphi$ in Eq.~\eqref{trans}, they become
\begin{align}
	&(\widetilde\varphi^2) = \widetilde\varphi_1^2-\widetilde\varphi_2^2+\widetilde\varphi_3^2,\\
	&(\widetilde\varphi^2)' = \widetilde\varphi_1^2+\frac{1}{2}(\widetilde\varphi_2^2+2\sqrt{3}\widetilde\varphi_2\widetilde\varphi_3-\widetilde\varphi_3^2),\\
	&(\widetilde\varphi^2)'' = \widetilde\varphi_1^2+\frac{1}{2}(\widetilde\varphi_2^2-2\sqrt{3}\widetilde\varphi_2\widetilde\varphi_3-\widetilde\varphi_3^2),\\
	&(\widetilde\varphi^2)_{\bf 3_S} = \begin{pmatrix}
		\widetilde\varphi_2^2+\widetilde\varphi_3^2 -2\sqrt{2}\widetilde\varphi_1\widetilde\varphi_3 \\
		\widetilde\varphi_2^2+\widetilde\varphi_3^2 +\sqrt{2}\widetilde\varphi_1(\sqrt{3}\widetilde\varphi_2+\widetilde\varphi_3) \\
		\widetilde\varphi_2^2+\widetilde\varphi_3^2 +\sqrt{2}\widetilde\varphi_1(-\sqrt{3}\widetilde\varphi_2+\widetilde\varphi_3)
	\end{pmatrix},~~~
	(\widetilde\varphi^2)_{\bf 3_A} = 0,\\
	&(\widetilde\varphi^{\,3})_{\bf 1}=3\sqrt{3}\widetilde\varphi_1 (\widetilde\varphi_2^2+\widetilde\varphi_3^2).
\end{align}



\end{document}